\documentclass[%
 reprint,
 amsmath,amssymb,
 aps,
prb,
]{revtex4-2}

\usepackage{graphicx}
\usepackage{dcolumn}
\usepackage{bm}

\begin{document}
\preprint{APS/123-QED}

\title{Driving force of atomic ordering in Fe$_{1-x}$Pt$_{x}$, investigated by density functional theory and machine-learning interatomic potentials Monte Carlo simulations}

\author{Tomoyuki Tsuyama}%
\email{tsuyama.tomoyuki.xixae@resonac.com}
\address{Resonac Corporation, Research Center for Computational Science and Informatics \\ 8, Ebisu-cho, Kanagawa-ku, Yokohama, Kanagawa. 221-8517, Japan}%
\author{Takeshi Kaneshita}%
\address{Resonac Corporation, Research Center for Computational Science and Informatics \\ 8, Ebisu-cho, Kanagawa-ku, Yokohama, Kanagawa. 221-8517, Japan}%
\author{Akira Matsui}%
\address{Resonac Corporation, Research Center for Computational Science and Informatics \\ 8, Ebisu-cho, Kanagawa-ku, Yokohama, Kanagawa. 221-8517, Japan}%
\author{Kohei Ochiai}%
\address{Resonac Corporation, Research Center for Computational Science and Informatics \\ 8, Ebisu-cho, Kanagawa-ku, Yokohama, Kanagawa. 221-8517, Japan}%
\author{Hiroaki Tanaka}%
\address{Resonac Corporation, Research Center for Computational Science and Informatics \\ 8, Ebisu-cho, Kanagawa-ku, Yokohama, Kanagawa. 221-8517, Japan}%
\author{Ryohei Kondo}%
\address{Resonac Corporation, Research Center for Computational Science and Informatics \\ 8, Ebisu-cho, Kanagawa-ku, Yokohama, Kanagawa. 221-8517, Japan}%
\author{Takayuki Fukushima}%
\address{Resonac Hard Disk Corporation, Research \& development Center \\ 5-1, Yawatakaigan-dori, Ichihara, Chiba, 290-0067, Japan}%
\author{Haruhisa Ohashi}%
\address{Resonac Hard Disk Corporation, Research \& development Center \\ 5-1, Yawatakaigan-dori, Ichihara, Chiba, 290-0067, Japan}%
\author{Atsushi Hashimoto}%
\email{hashimoto.atsushi.xhtcs@resonac.com}
\address{Resonac Hard Disk Corporation, Research \& development Center \\ 5-1, Yawatakaigan-dori, Ichihara, Chiba, 290-0067, Japan}%
\author{Yoshishige Okuno}%
\address{Resonac Corporation, Research Center for Computational Science and Informatics \\ 8, Ebisu-cho, Kanagawa-ku, Yokohama, Kanagawa. 221-8517, Japan}%
\author{Jian-Gang Zhu}%
\address{Data Storage Systems Center, Carnegie Mellon University, Pittsburgh, PA 15213, USA.}%
\address{Electrical and Computer Engineering Department, Carnegie Mellon University, Pittsburgh, PA 15213, USA.}%
\address{Materials Science and Engineering Department, Carnegie Mellon University, Pittsburgh, PA 15213, USA.}%

\date{\today}

\begin{abstract}
  We report the mechanisms of atomic ordering in Fe$_{1-x}$Pt$_{x}$ alloys using density functional theory (DFT) and machine-learning interatomic potential Monte Carlo (MLIP-MC) simulations.  
  We clarified that the formation enthalpy of the ordered phase was significantly enhanced by spin polarization compared to that of the disordered phase.  
  Analysis of the density of states indicated that coherence in local potentials in the ordered phase brings energy gain over the disordered phases, when spin is considered.  
  MLIP-MC simulations were performed to investigate the phase transition of atomic ordering at a finite temperature.
  The model trained using the DFT dataset with spin polarization exhibited quantitatively good agreement with previous experiments and thermodynamic calculations across a wide range of Pt compositions, whereas the model without spin significantly underestimated the transition temperature.  
  Through this study, we clarified that spin polarization is essential for accurately accounting for the ordered phase in Fe-Pt bimetallic alloys, even above the Curie temperature, possibly because of the remaining short-range spin order.  
\end{abstract}
\maketitle




\section{\label{sec:intro}Introduction}
  FePt has garnered significant attention in both industrial applications and fundamental research, particularly when the crystal adopts the $L1_{0}$ (ordered) phase, which features alternating pure planes of Fe and Pt along the [001] direction \cite{FePt-science, FePt-IEEE, FePt-AIP, FePt-JPSJ}.  
  This highly anisotropic crystal structure, characterized by substantial spin-orbit coupling because of the high atomic number of Pt, exhibits remarkable perpendicular magnetic anisotropy.  
  This anisotropy enables the magnetization to remain stable by mitigating thermal fluctuations and other noise sources.  
Consequently, FePt is a primary candidate material for various magnetic recording technologies, such as heat-assisted magnetic recording for hard disk drives (HAMR) \cite{HAMR-roadmap, FePt-HAMR1, FePt-HAMR2} and magnetic random access memory (MRAM) \cite{FePt-MRAM}.  

  Given that the magnitude of perpendicular magnetic anisotropy is influenced by the degree of ordering in the $L1_{0}$ phase, it is crucial to understand the underlying mechanisms of the stability associated with the atomic order in FePt alloys \cite{JZhang2013}.  

  The atomic ordering of bimetallic compounds has been systematically studied \cite{SRO_bimetal}.  
  In certain bimetallic alloy systems, such as FeCo, theoretical studies have demonstrated that magnetism contributes to the stabilization of atomic order \cite{FeCo-PRB,FeCo2-PRB}.  
  The effects of spins on the CoPt ground state were investigated using density functional theory (DFT), concluding that magnetism enhances the stability of the $L1_{0}$ phase through $d$-band filling \cite{Karoui_2013}.  
  The authors posited that a similar theory can be applied to FePt.  
  However, the phase stability of $L1_{0}$ compared to that of the $A1$ (disordered) phase, which is of primary interest concerning atomic ordering in the system, was not addressed.
  Furthermore, the actual impact of spin polarization at finite temperatures, particularly near the order-to-disorder phase transitions, has not been verified.  

  In this study, we comprehensively discuss the impact of spin polarization on the ordered phase in both ground states and finite temperatures.
  The formation energy of Fe$_{1-x}$Pt$_{x}$ was obtained by DFT and machine-learning interatomic potentials (MLIPs), concluding that spin polarization plays an important role in amplifying the stability of ordered phases rather than disordered phases or phase separations.  
  The density of states (DOS) of FePt indicates that the coherence of potentials in the ordered phase with spin polarization provides stability over disordered phases.  
  The order-to-disorder phase transition of Fe$_{1-x}$Pt$_{x}$ was simulated using machine-learning interatomic potential Monte Carlo (MLIP-MC), which reproduced previous studies quantitatively by considering spin polarization in the DFT dataset.  
  Through these methods, we directly showed that spin polarization is the critical driver in the form of ordered phases of Fe$_{1-x}$Pt$_{x}$ even above Curie temperature.

\section{\label{sec:calc_condition}Methods}
\subsection{Density Functional Theory Calculations}
We performed DFT simulations using the Vienna \textit{ab initio} Simulation Package (VASP), version 6.3.2, with Perdew-Burke-Ernzerhof generalized-gradient approximation exchange-correlation functionals \cite{VASP, VASP2, PBE, GGA1, GGA2, GGA3}.  
The cutoff energy was set to 295 eV, which is 10\% higher than the recommended value for Fe and Pt pseudopotentials, which consider the $3d^{7}4s^{1}$ states for Fe and $5d^{9}4s^{1}$ states for Pt as valence electrons, respectively.  
Monkhorst-Pack grids of 20 $\times$ 20 $\times$ 20 $k$-point mesh were employed except for the DOS analysis of the disordered phase \cite{Monkhorst}.  
For the DOS analysis of the disordered phase, we employed a supercell approach composed of Fe$_{54}$Pt$_{54}$ with randomized atomic order, calculated using Monkhorst-Pack grids of 6 $\times$ 6 $\times$ 6 $k$-point mesh \cite{CPA, CPA2}.  
To ensure that the sampled structures represent the disordered phase, we generated five distinct randomized ordered structures and conducted calculations.  
We ensured that the energy and band structures could be sufficiently converged using the supercell approach.  

\subsection{Machine-Learning Interatomic Potentials}
We employed MLIPs to evaluate the energy of disordered states in Fe-Pt alloy systems and to conduct on-lattice Monte Carlo (MC) simulations for the atomic ordering phase at finite temperatures because of the following two unique characteristics.  
First, MLIPs offer significant advantages in terms of computational cost over DFT while maintaining sufficiently high energy accuracy, which makes it one of the most suitable methods to perform large-scale randomized phase simulations \cite{Behler_2016, Bartok_2013}.  
Second, because MLIPs can be constructed purely from DFT data, it is possible to simulate virtual physical states, such as the non-magnetic phase of Fe-Pt alloy systems considered in this study, derived from appropriate DFT settings.  

Among the many MLIP schemes, recent benchmarks have demonstrated that moment tensor potentials offer competent balances between accuracy and computational efficiency \cite{MLIP_bench, MLIP1, MLIP2}.  
Thus, we selected moment tensor potentials as the MLIPs, which were trained using the following procedures.  

Molecular dynamics (MD) simulations using a Large-scale Atomic/Molecular Massively Parallel Simulator (LAMMPS) were performed with interatomic potentials derived from the modified embedded atom method to generate atomic structures for training \cite{LAMMPS, FePt_meam}.  
The repeated structures of 3 $\times$ 3 $\times$ 3 of $L1_{0}$ FePt, $L1_{2}$ FePt$_{3}$, Fe$_{3}$Pt, pure bcc Fe, and fcc Pt were prepared as initial structures for the bulk.  
(111) and (100) surface slabs were also prepared to represent the slab structures.  
MD simulations at various temperatures ($T$ = 500, 1000, 1500, 2000, and 2500 K) were conducted under NPT for bulk and NVT for slabs, executed for 50 picoseconds with a time step of 2 femtoseconds.  
During the MD simulations, pairs of nearest neighbor Fe and Pt atoms were randomly selected, and their atomic positions were exchanged every 2 picoseconds to expand the chemical space and to cover disordered phases, which cannot be achieved solely through the timescale of MD simulations because of high energy barriers. 

We employed dimensionality reduction and stratified sampling encoded by the Materials 3-body Graph Network (M3G-Net) universal model, to efficiently cover the extensive structural and chemical space of the training dataset for the materials of interest\cite{direct_sampling, m3gnet}. 
We performed DFT calculations using VASP under the aforementioned conditions for the 1,000 structures selected using this method.  
The labeled dataset was partitioned into 900 training and 100 validation datasets.  
Training was conducted using a cutoff radius of 0.8 nm.  
The root mean square error in energy was achieved 4 meV/atom in the validation data, which is sufficiently small to facilitate discussions on the formation enthalpy of Fe$_{1-x}$Pt$_{x}$ systems \cite{FePt_formation_eng}.   

\subsection{Monte Carlo Simulations}
On-lattice MC simulations using the Metropolis method were performed to simulate the transition temperature of the Fe-Pt binary system, which has been extensively applied to describe phase transitions of atomic ordering \cite{FePt_MC1, FePt_MC2, CoPt_MC}.    

In this study, we evaluated the energy of the alloy using MLIPs from the atomic coordinates updated by MC. 
The degree of atomic order during the simulations was quantized using the long-range order (LRO) parameter, defined in references \cite{FePt_MC1, long_range_order_parameter}.  

We prepared a 14 $\times$ 14 $\times$ 14 supercell of Fe$_{1-x}$Pt$_{x}$, comprising 10,976 atoms with periodic boundary conditions.  
To simulate the order-to-disorder phase transitions, the temperautre was gradually increased by 20 K.  
In total, 100,000 MC steps were performed at a single temperature.  

To compare the transition temperatures simulated by MLIP-MC with reference, thermodynamic calculations were conducted using the CALPHAD (CALculation of PHAse Diagrams, Computer Coupling of Phase Diagrams and Thermochemistry) method with Thermo-Calc, which reproduces the experimental results \cite{ThermoCalc, FePt_phase_diagram_experiment}. 
The calculations employed the thermodynamic database TCNOBL3.

\section{\label{sec:results}Results and discussion}
\subsection{Formation Energy}
In this subsection we discuss the formation energies of Fe$_{1-x}$Pt$_{x}$ alloys using DFT and MLIPs with and without spin, showing the critical role of spins on ordered states in Fe-Pt alloy systems.  

We obtained the formation enthalpy ($\Delta H$, at $T = 0$ K) of Fe$_{1-x}$Pt$_{x}$ alloys, defined as  
$$
\Delta H = \left( E^{\left( \mathrm{Fe}_{m} \mathrm{Pt}_{n} \right)} - mE^{\mathrm{Fe}} - nE^{\mathrm{Pt}} \right) / (m+n),
$$
where $E^{\left( \mathrm{Fe}_{m} \mathrm{Pt}_{n} \right)}$ denotes the total energy of mixed Fe and Pt systems composed of $m$ Fe and $n$ Pt atoms, respectively. 
$E^{\mathrm{Fe}}$ and $E^{\mathrm{Pt}}$ represent the energy per atom of pure Fe and Pt metals in fcc structures, because of the simpler discussions on the formation energy of bimetallic systems in Fe-Pt \cite{FePt_formation_eng}.

\begin{figure}[tbp]
  \begin{center}
    \includegraphics[width=3.37in]{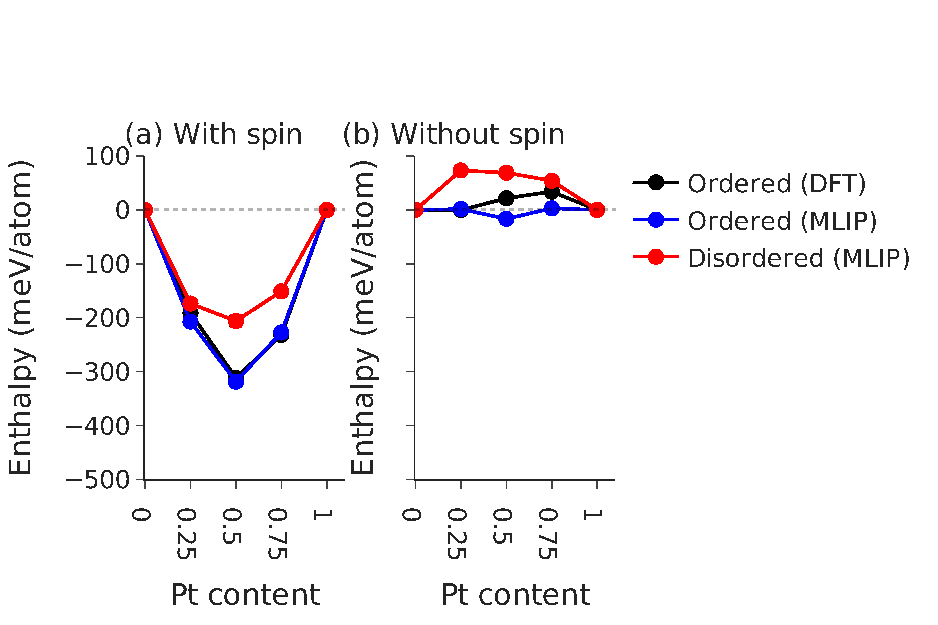}
    \caption{\label{fig:formation_energy} The formation energy of Fe$_{1-x}$Pt$_{x}$ for various Pt contents calculated by DFT and MLIPs considering spin polarization (a) and without spin polarization (b). Ordered phases ($L1_{0}$ and $L1_{2}$) were calculated by both DFT and MLIPs, and disordered phases were calculated only by MLIPs. The energies of the disordered phase are sampled from 100 structures and plotted with error bars, which are difficult to discern because of their small size.}
  \end{center}
\end{figure}  

Figure \ref{fig:formation_energy} illustrates the formation energies of Fe$_{1-x}$Pt$_{x}$ alloy systems with various Pt contents calculated by DFT and MLIPs with spin (a), and without spin (b) polarizations. 
We compare the formation energy obtained by MLIPs and DFT on ordered phases to confirm the accuracy of the MLIP models, and both methods exhibit good quantitative agreement with previous studies \cite{FePt_formation_eng}.  
Note that we did not take the antiferromagnetic state into account for FePt$_{3}$ compositions, which have a slightly larger formation energy than ferromagnetic states, because the energy difference between these phases is relatively small compared to the energies discussed in this paper.  

Figure \ref{fig:formation_energy} (a) shows the stably ordered phases of FePt and FePt$_{3}$, and Fe$_{3}$Pt.  
These ordered phases are more stable than the disordered phases calculated using MLIPs for all compositions when considering spin polarizations.  

Figure \ref{fig:formation_energy} (b) suggests that the energy gains obtained by mixing Fe and Pt almost vanish or even become unstable without considering spin polarizations in ordered phases in the entire composition.  
The disordered phases obtained by MLIPs were even more unstable and could lead to phase separation.  
The energy differences between the ordered and disordered phases were smaller without spin polarization, which could be a factor in decreasing the order-to-disorder phase transition temperature.  

In this subsection, we computed the formation energy of Fe$_{1-x}$Pt$_{x}$ alloys, clarifying the importance of spin polarization in stabilizing the atomic ordered phases in Fe-Pt bimetallic systems.

\subsection{Density of States}
\begin{figure}[tbp]
  \begin{center}
    \includegraphics[width=3.37in]{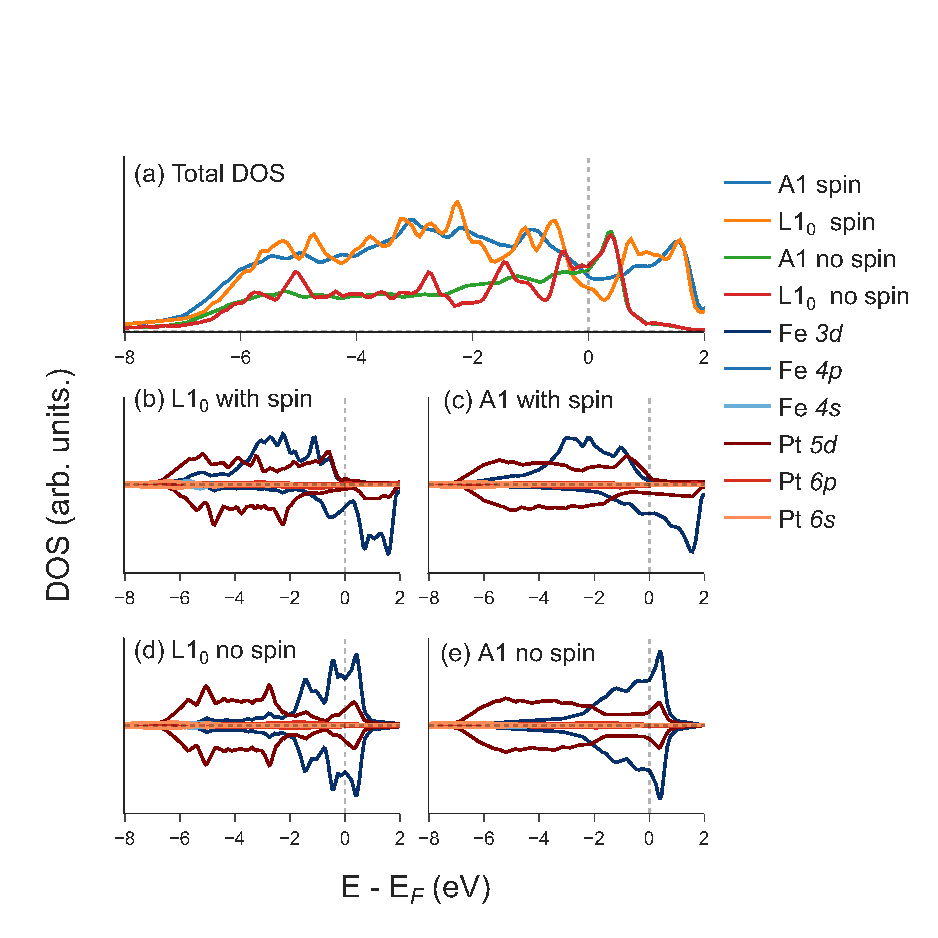}
    \caption{\label{fig:DOS} The total density of states (DOS) and projected density of states (pDOS) comparing the $L1_{0}$ and $A1$ phases. The calculations were done with and without spin polarization for each phase. Panel (a) shows the total DOS for these states. The pDOS, categorized by element, orbitals, and spins, is plotted for spin-polarized $L1_{0}$ (b), $A1$ (c), non-spin-polarized $L1_{0}$ (d), and $A1$ (e),  where positive and negative values represent spin-up and spin-down DOS, respectively.}
  \end{center}
\end{figure}

  We analyzed the DOS of FePt to determine the origin of the stability in the ordered phases enhanced by spin.  

  In Figure \ref{fig:DOS} (a), the total DOS is compared between $L1_{0}$ and $A1$ phases with and without spin polarization.  
  We observe two tendencies: spin polarization modulates the peak positions, and ordering ($L1_{0}$ and $A1$ phases) influences the smoothness of the DOS.  
  The modulation of the peak position by the spins can be explained by the spin-up and spin-down splits of the DOS.  
  The sharpness of the DOS in $L1_{0}$ results from the coherence of the potentials because of atomic ordering \cite{FePt_do_DOS}.  

  The atom-projected electronic DOS for each orbital is displayed for spin-polarized $L1_{0}$ in (b), $A1$ in (c), non-spin-polarized $L1_{0}$ in (d), and $A1$ in (e).

  The non-spin-polarized DOS exhibits high values near $E_{F}$.  
  Consequently, the asymmetric spin-split state gains energy by reducing the DOS value at $E_{F}$ (i.e., satisfying the Stoner criterion).  
  Thus, the ferromagnetic state was preferred as the ground state for both the $L1_{0}$ and $A1$ phases.  
  The reduction of the DOS value at $E_{F}$ can be observed in $L1_{0}$ compared to $A1$ phase in spin polarized results, which is likely advantageous for the $L1_{0}$ phase stabilization, as it shifts the center of the occupied state toward lower energy ($\int_{-\infty}^{E_F} \epsilon D(\epsilon) \, d\epsilon$).  
  By comparing Figure \ref{fig:DOS} (b) and (c), the modulations of the DOS value at $E_{F}$ are mainly attributed to Fe $3d$ orbitals and Pt $5d$ states originating from the coherence of localized potentials \cite{FePt_do_DOS}.  
  Ueda \textit{et al}. compared the valence-band hard-x-ray photoemission spectra of FePt $L1_{0}$ and $A1$ phases, showing higher intensity at $E_{F}$ in the $A1$ phase, which qualitatively agrees with our DOS analysis \cite{FePt_XPS}.  

  In contrast, without spin polarization, the changes in the DOS value at $E_{F}$ between the $L1_{0}$ and $A1$ phases were subtle.  
  Thus, the stability of $L1_{0}$ compared with that of $A1$ without spin polarization is expected to be less significant than that with spin polarization.

  Another aspect explaining the stability of $L1_{0}$ owing to spin polarization is the $d$-band filling effect, which has been reported in several previous studies \cite{SRO_bimetal, Karoui_2013}.  
  The changes in the occupation of the $d$-orbitals brought about by spin polarization elucidate the stability of the $L1_{0}$ phases, as mediated by pair interactions.  

  To summarize the results of the DOS analysis, the energy stability of ordered phases over disordered ones is enhanced by spin polarization, as the ordered state has a sharp DOS due to coherence in local potentials which results in lowering the total energy below $E_{F}$.  

\subsection{Order-to-Disorder Phase Transition}
\begin{figure}[tbp]
  \begin{center}
    \includegraphics[width=3.37in]{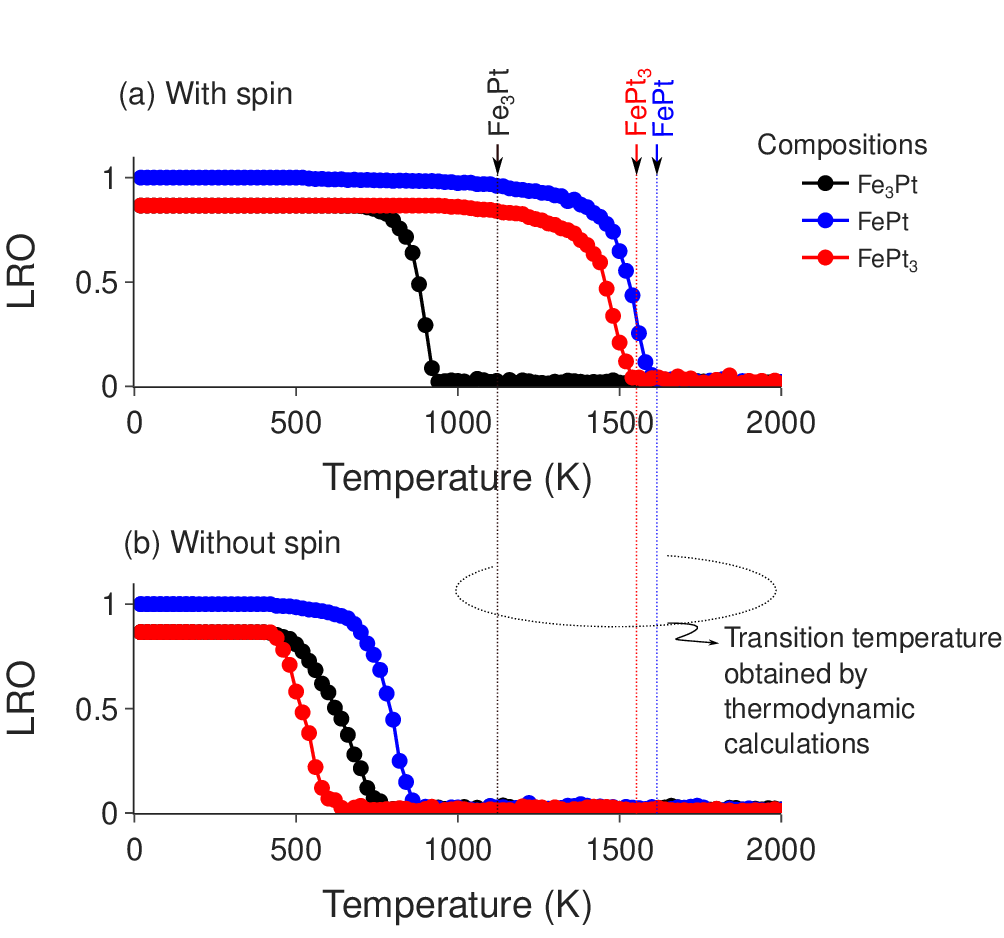}
    \caption{\label{fig:phase_transitions} Evolutions of the norm of long-range order (LRO) parameters as defined in reference \cite{FePt_MC1, long_range_order_parameter}, simulated by MLIP-MC for FePt, FePt$_{3}$, and Fe$_{3}$Pt. The energy is evaluated using MLIPs trained by the datasets with (a) and without (b) spin polarization. The vertical dotted-lines indicate the transition temperatures obtained from thermodynamic simulations.}
  \end{center}
\end{figure}

  In this subsection, we demonstrate that the MLIP-MC trained by the DFT dataset with spin polarization reproduces the phase diagram of the Fe-Pt binary system acquired by thermodynamic calculations, whereas the one trained without spin polarizations tremendously underestimates the transition temperatures.  

  Figure \ref{fig:phase_transitions} shows the temperature dependence of the norm of LRO parameters (i.e., magnitude of LRO vectors) for FePt, FePt$_{3}$, and Fe$_{3}$Pt obtained by the MLIP-MC simulations with (a) and without (b) spin polarization in the training datasets.  
  The order-to-disorder transition temperatures simulated by the DFT-trained model with spin polarization are in qualitative agreement with experiments within $\sim$ 200 K.  
  However, without spin polarization shown in panel (b), the transition temperatures are underestimated by more than $\sim$ 800 K in FePt.  

\begin{figure}[tbp]
  \begin{center}
    \includegraphics[width=3.37in]{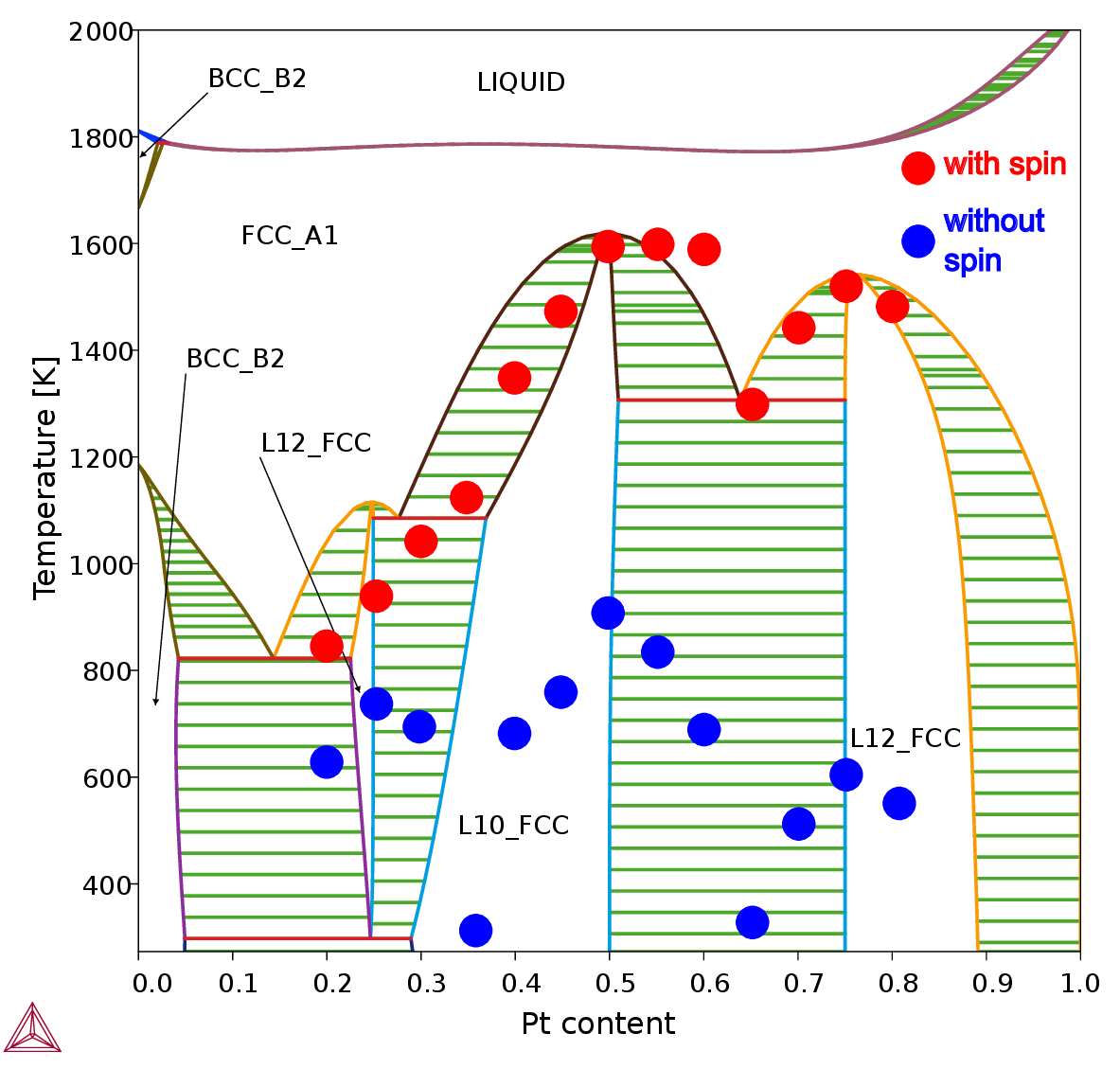}
    \caption{\label{fig:phase_diagram} The order-to-disorder phase diagram of Fe${1-x}$Pt${x}$ is presented, comparing the results obtained from thermodynamic (CALPHAD) calculations with those from MLIP-MC simulations. The red circle plots represent the transition temperatures simulated by MLIP-MC, which was trained using a DFT dataset that incorporates spin, while the blue circle plots correspond to the transition temperatures obtained without considering spin.}
  \end{center}
\end{figure}

  We also performed MC simulations on non-stoichiometric compositions by randomly substituting Fe or Pt in FePt, FePt$_{3}$, or Fe$_{3}$Pt (replacing atoms from the nearest stoichiometric compositions) to achieve the desired compositions.  
  In Figure \ref{fig:phase_diagram}, we mapped the transition temperatures obtained from the MC simulations onto the phase diagram derived from the thermodynamic (CALPHAD) calculations.
  The transition temperatures from the MLIP-MC trained by the spin-polarized DFT dataset (red circles) show good agreement with those from thermodynamic calculations, although Fe-rich compositions tend be underestimated by up to $\sim$ 200 K. 
  Previous MC studies of Fe-Pt alloy systems, which considered up to the $2^{nd}$ nearest neighbor pair-interactions and the modified embedded atom method, overestimated the transition temperature of Fe$_{3}$Pt by $\sim$ 400 K \cite{FePt_MC1, FePt_meam}.  

  In contrast, MLIP-MC using the non-spin-polarized DFT dataset (blue circles) significantly underestimated the transition temperature.  
  These results directly clarify that spin polarization is crucial for quantitatively reproducing the order-to-disorder phase transition temperature acquired by thermodynamic simulations and experiments \cite{FePt_phase_diagram_experiment}.  
  This is not obvious, as the order-to-disorder phase transition temperatures in these alloys are above the Curie temperature, and the macroscopic magnetism should vanish.  
  We deduced that short-range spin interactions persist and are responsible for the stabilization of ordered states up to the order-to-disorder phase transition temperature.

  The importance of spin in the ground state of CoPt alloy systems has been demonstrated using DFT \cite{Karoui_2013}. 
  Neutron diffraction studies have demonstrated that pair interactions are not changed by the Curie temperature, and that order-to-disorder transition temperature can be accurately simulated by MC, whose parameter is determined by neutron diffraction measurements above the Curie temperature \cite{CoPt_MC, CoPt_Neutron}.  
  These facts imply that spin polarization has an important influences on pair interactions even above the Curie temperature.  
  Several previous studies have shown that the short-range spin order persists above the Curie temperature \cite{SRO_Fe, SRO}.  
  Leroux \textit{et al}. suggested that the short-range magnetic order above the Curie temperature of CoPt causes a unique temperature dependence of the resistivity at order-to-disorder phase transitions \cite{SRO_CoPt}.  
  The short-range spin order persists above the Curie temperature at not less than 85\% of the $T$ = 0 when comparing the experimental and calculation results for Fe$_{3}$Pt \cite{SRO_Fe3Pt}.  
  From our results and those of previous studies on systems similar to FePt, we deduce that local spin order still exists and contributes to the local stability of the ordered phase.  
  Further verification and detailed mechanic studies on the stabilization of the ordered phase in Fe$_{1-x}$Pt$_{x}$ by spin polarization above the Curie temperature remain future work.  

  In this subsection, we demonstrate that it is necessary to consider spin polarization to reproduce the experimental phase diagram for the order-to-disorder transition of Fe$_{1-x}$Pt$_{x}$ systems through MLIP-MC simulations at finite temperatures.

\section{\label{sec:conclusion}Conclusion}
In summary, we investigated the underlying physics of ordered phases of Fe$_{1-x}$Pt$_{x}$ alloys using DFT and MLIPs.  

  The formation energies calculated by DFT and MLIPs revealed that spin polarization is essential for explaining the phase stability of ordered Fe$_{1-x}$Pt$_{x}$.  
  The energy gains of transitioning from the disordered phase to the ordered phase are also enhanced by considering spin.  
 
  The higher DOS value in the $A1$ phase at $E_{F}$ compared to the $L1_{0}$ phase in spin-polarized systems indicates instability in the $A1$ phase caused by the local potential fluctuations in randomized atomic order.  
  In contrast, modulations in the DOS at $E_{F}$ without spin polarization are less affected because of the lack of spin splitting: consequently the stabilization of $L1_{0}$ compared to $A1$ phase is weaker without spin polarization. 

  MLIP-MC trained by the DFT dataset considering spins demonstrated quantitatively good agreement in the order-to-disorder phase transition temperatures with thermo dynamic calculations and previous experiments throughout the entire composition of Fe$_{1-x}$Pt$_{1-x}$.  

  From the above results, we conclude that spin polarization is essential for characterizing the ordered phases of Fe-Pt bimetallic systems.  
  Investigating the origin of the necessity for spin polarization above the Curie temperature will be the focus of future work, but the existence of short-range spin order may be one possible explanation for this phenomenon.


\nocite{*}
\bibliography{reference_v01}

\end{document}